\documentclass[aps,prb,twocolumn,amsmath,amssymb,superscriptaddress,floatfix]{revtex4}
\usepackage{graphicx}
\usepackage{bm}
\usepackage{afterpage}
\usepackage[usenames]{color}
\bibstyle{apsrev.bib}

\newcommand{\be}{\begin{equation}}
\newcommand{\ee}{\end{equation}}
\newcommand{\beqn}{\begin{eqnarray}}
\newcommand{\eeqn}{\end{eqnarray}}

\begin{document}

\title{Quantum multicritical point in the two- and three-dimensional random transverse-field Ising model}

\author{Istv\'an A. Kov\'acs}
\email{istvan.kovacs@northwestern.edu}
\affiliation{Department of Physics and Astronomy, Northwestern University, Evanston, IL 60208}
\affiliation{Northwestern Institute on Complex Systems, Northwestern University, Evanston, IL 60208}
\date{\today}

\begin{abstract}
Quantum multicritical points (QMCPs) emerge at the junction of two or more quantum phase transitions due to the interplay of disparate fluctuations, leading to novel universality classes. While quantum critical points have been well characterized, our understanding of QMCPs is much more limited, even though they might be less elusive to study experimentally than quantum critical points.
Here, we characterize the QMCP of an interacting heterogeneous quantum system in two and three dimensions, the ferromagnetic random transverse-field Ising model (RTIM). The QMCP of the RTIM emerges due to both geometric and quantum fluctuations, studied here numerically by 
the strong disorder renormalization group method. The QMCP of the RTIM is found to exhibit ultraslow, activated dynamic scaling, governed by an infinite disorder fixed point. This ensures that the obtained multicritical exponents tend to the exact values at large scales, while also being universal --- i.e. independent of the form of disorder ---, providing a solid theoretical basis for future experiments.
\end{abstract}


\maketitle
\section{Introduction}

\begin{figure}[ht]
\begin{center}
\includegraphics[width=2.2in,angle=0]{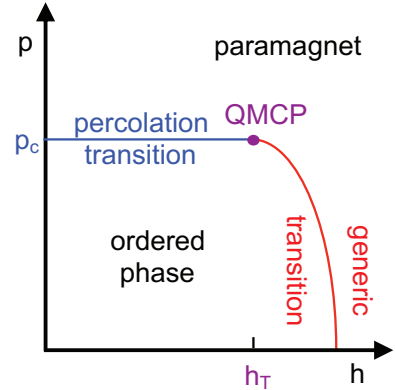}
\end{center}
\vskip -0.75cm
\caption{
\label{fig_1} (Color online) \textbf{Phase diagram of the RTIM:}
In two and higher dimensions the phase transition is either driven by geometric or quantum fluctuations. Quantum critical behavior at the bond dilution $p=p_c$ (blue line) is controlled by the percolation fixed point\cite{senthil_sachdev}, while below $p_c$ the generic 'quantum percolation' universality class is found (red), by tuning the $h$ magnetic field. Both the percolation and generic universality classes are known examples of activated dynamic scaling, see text. In this paper, we characterize the quantum multicritical point (QMCP, purple), and show that it is marked by activated dynamic scaling, with a new set of universal exponents. 
}
\end{figure}

Understanding emergent, collective phenomena in interacting quantum systems is among the fundamental problems of modern physics, with applications in solid state physics, quantum field-theory, quantum information and statistical mechanics\cite{sachdev}. As prominent examples, quantum phase transitions play an important role in rare-earth magnetic insulators\cite{BRA96}, heavy-fermion compounds \cite{HeavyF,Lohn96}, high-temperature superconductors\cite{hightc,sciencereview} and two-dimensional electron gases \cite{sondhi,KMBFPD95}, among others. These transitions take place in the ground state of the quantum system by varying the strength of quantum fluctuations through a control parameter, such as a transverse field, or the dilution fraction of magnetic vs.~non-magnetic atoms. Experimentally, at least in ferromagnetic systems, quantum critical points (QCPs) are hard to access as they might either change to first-order transition or get buried inside an intervening phase \cite{buried1,buried2, buried3, buried4}.  As an alternative, quantum multicritical points (QMCPs) have been proposed as key concepts to understand the onset of the ordered phase in a range of quantum systems, arising naturally at the intersection of two or more quantum phase transitions (Fig.~\ref{fig_1}). 
For example, a ferromagnetic QMCP has been studied experimentally recently in the disordered compound $\mathrm{Nb}_{1-y}\mathrm{F}_{2+y}$ \cite{friedemann} to investigate the onset of long-range order. 
%
The way how such quenched disorder influences the properties of quantum phases and phase transitions is a principal theoretical question. Addressing it, the random transverse-field Ising model (RTIM) has been a paradigmatic model, given by the Hamiltonian
\be
{\cal H} =
-\sum_{\langle ij \rangle} J_{ij}\sigma_i^x \sigma_{j}^x-\sum_{i} h_i \sigma_i^z\;.
\label{eq:H}
\ee
Here, the $\sigma_i^{x,z}$ are Pauli-matrices at each site, while $\langle ij \rangle$ indicate pairs of sites of a diluted lattice, selected with bond probability $p$. The $J_{ij}>0$ couplings and the $h_i>0$ transverse fields are independent random numbers, taken from the distributions, $p(J)$ and $q(h)$, respectively. 

Detailed theoretical results about the generic QCP of the RTIM are known in one dimension (1D) due to a complete analytical solution of the strong disorder renormalization group (SDRG) treatment\cite{fisher}. A key observation is that the critical properties of the RTIM are governed by an infinite disorder fixed point (IDFP), in which the strength of disorder grows without limit during renormalization\cite{danielreview}. Therefore, the SDRG results are expected to be asymptotically exact in the vicinity of the critical point, which is indeed demonstrated by a comparison with independent analytical\cite{mccoywu,shankar} and numerical\cite{young_rieger96,bigpaper} works. 
The IDFP scenario has been shown also for the 2D and higher dimensional RTIM in numerical SDRG studies\cite{motrunich00,lin00,karevski01,lin07,yu07,2dRG,ddRG} as well as in 2D and 3D Monte Carlo simulations\cite{pich98, vojta09, vojta3D}. Besides the RTIM, well-known examples of IDFP are the random antiferromagnetic (AF) Heisenberg chain \cite{fisher94}, random quantum Potts\cite{senthil_majumdar} and Ashkin-Teller\cite{carlon-clock-at} models, and also non-equilibrium classical models, such as the random contact process \cite{hiv,vojta05}.

As illustrated in Fig.~\ref{fig_1}, in higher dimensions, the RTIM also undergoes a percolation QCP by tuning the bond percolation probability $p$, for sufficiently weak external fields\cite{2qcp, senthil_sachdev}. This percolation transition happens at the well known classical bond percolation critical point $p_c$, independently from the strength of the $h$ external field. Along this percolation line the ground state of ${\cal H}$ is given by a set of independently ordered clusters, which are in the same form as for percolation. The critical exponents are known exactly in 2D and to high precision in 3D as summarized in Table~\ref{table:1}. For $p>p_c$, at least one giant percolating cluster is present in the system, providing the basis of a generic quantum phase transition by tuning the external field $h$ to its critical value (which depends on $p$). 
Both along the percolation and the generic QCPs, the IDFP scenario leads to anomalous, activated dynamic scaling. 

At a critical point, as the system size increases, the characteristic time scale of the dynamics $\tau\sim 1/\epsilon$ becomes slower and slower, where $\epsilon$ stands for the characteristic energy scale of the sample, i.e.~the smallest energy gap. At conventional critical points, dynamical scaling is characterized by the $z$ dynamical exponent as $\epsilon~\sim L^{-z}$, relating the $\epsilon$ energy scale to the $L$ length scale of the system. 
As an extreme limit, activated dynamic scaling is manifested in the relation
\be
\ln \epsilon~\sim L^{\psi}\;,
\label{active}
\ee
formally corresponding to a diverging $z$ dynamical exponent, characteristic of ultraslow dynamics, a hallmark of IDFPs. 
%
Both the percolation and generic QCPs have been studied either analytically or numerically to high precision in 2 and 3D, including the quantum entanglement properties\cite{refael,lin07,yu07, kovacs, kovacs3dperc, kovacs_igloi12}.
The last missing piece from a complete understanding of the RTIM phase diagram in Fig.~\ref{fig_1} is the behavior at the QMCP. 
To fill this gap, we extend the investigations about the critical behavior of the RTIM into the QMCP, where both geometric and quantum fluctuations diverge. 

The rest of paper is organized as follows. The SDRG method is described in Sec.\ref{Sec:SDRG}. Results about the multicritical parameters are calculated in Sec.\ref{Sec:critical} and discussed in Sec.\ref{Sec:discussion}.

\begin{figure}[ht]
\begin{center}
\includegraphics[width=3.4in,angle=0]{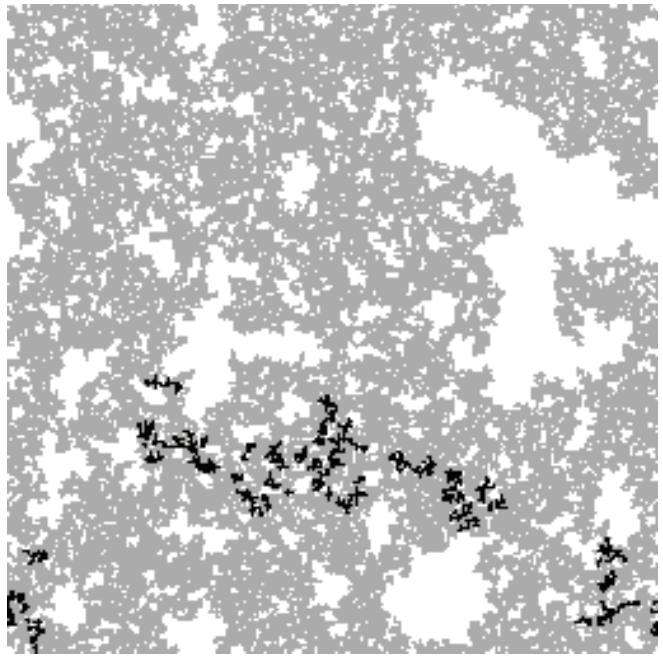}
\end{center}
\vskip -.5cm
\caption{
\label{fig_2} 
\textbf{Ground state at the QMCP:}
Illustration of the correlation cluster (black) at the QMCP for $L=256$ in two dimensions with box-$h$ disorder. The encompassing critical bond percolation cluster is indicated in gray. In stark contrast to percolation (gray), the multicritical correlation cluster (black) is a disconnected fractal, similarly to those at the generic QCP\cite{kovacs_igloi12,clusters}, although characterized by a new value of the fractal dimension.
}
\end{figure}


\section{SDRG procedure}
\label{Sec:SDRG}

The SDRG approach\cite{im} has been a powerful way to study the critical behavior of disordered quantum systems, introduced by Ma, Dasgupta and Hu\cite{mdh} and further developed by D. Fisher\cite{fisher} and others. 
In the SDRG, an effective low-energy description is created successively, eliminating at each step the largest local term in the Hamiltonian, continuously lowering the energy scale. 
Depending on whether the largest local term is a coupling or a transverse field, new terms are generated between remaining sites by second-order perturbation method. When the largest term is a coupling, $J_{ij}$, the two connected spins merge into a spin cluster of the joint moment, $\tilde{\mu}=\mu_i+\mu_j$, which is placed in an effective transverse field of strength $\tilde{h}=h_i h_j/J_{ij}$. 
On the other hand, when the largest term is transverse field, $h_i$, the spin is eliminated and new effective couplings are generated between each pair of neighboring spins, say $j$ and $k$, having a value $\tilde{J}_{jk}=J_{ji}J_{ik}/h_i$. 
As a result, clusters of sites are successively created and eliminated, yielding a ground state decomposition in terms of independent spin clusters. 
Then, the characteristic energy scale is proportional to the smallest (effective) transverse field of any of the clusters, $\epsilon(s)$, in a given sample, $s$.
The magnetization cluster of the sample is the cluster corresponding to the smallest energy gap.
 
If, at any step, two parallel couplings appear between the same spins, the maximum of them is taken. 
Application of this 'maximum rule' is exact at an IDFP (where the distribution of the couplings becomes extremely broad) and leads to a highly efficient SDRG algorithm\cite{ddRG}. Our implementation of which needs only $t \sim \mathcal{O}(N \log N + E)$ CPU time and $\mathcal{O}(E)$ memory to renormalize a system with $N$ sites and $E$ interactions.

\section{Calculation of multicritical parameters}
\label{Sec:critical}

To investigate the universality of the results we have used two different standard forms of disorder\cite{2dRG, ddRG}. In both cases, the couplings are uniformly distributed in $(0,1]$. For 'box-$h$' disorder also the transverse fields are uniformly distributed in $(0,h]$, whereas for 'fixed-$h$' disorder we have a constant transverse field $h_i=h,\, \forall i$. We used the logarithmic variable, $\theta=\ln(h)$, as a quantum control parameter. We studied the QMCP in two and three dimensions, on cubic lattices with periodic boundary conditions up to over a million sites. 
The bond probability was set to its critical value $p_c(\mathrm{2D})=0.5$ and $p_c(\mathrm{3D})=0.248812$\cite{3dperc}.
%
The number of realizations used in the calculations were typically $10^5$, with at least $10^4$ samples for the largest sizes.


\subsection{Locating the Multicritical Point}


\begin{figure}[!ht]
\begin{center}
\includegraphics[width=3.3in,angle=0]{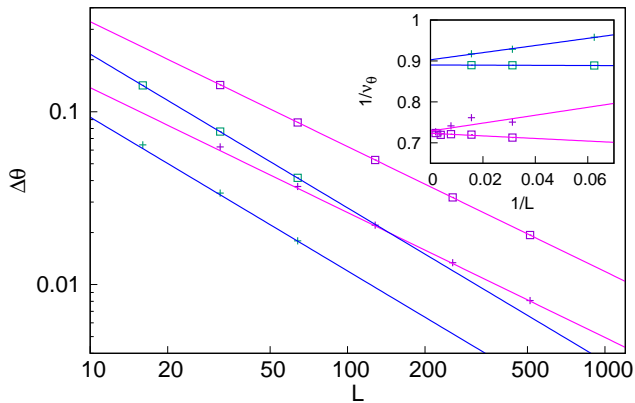}
\end{center}
\vskip -.75cm
\caption{
\label{fig_3} (Color online) \textbf{Determining the correlation length exponent:}
 Width of the distribution of the pseudo-multicritical points $\theta$ as a function of the length of the system in a log-log plot
for the 2D (purple, magenta) and 3D (green, blue) models for two types of disorder (box-$h$: $\boxdot$, fixed-$h$: $+$). The slope of the straight lines is given by $1/\nu_\theta$, where the extrapolated $\nu_\theta$ correlation length exponents are listed in Table~\ref{table:1}. The error of the data points is smaller than the symbol size.
Inset:  Two-point estimates of $1/\nu_\theta$ as the function of the inverse length, as well as the linear extrapolations, leading to the values in Table~\ref{table:1}. 
}
\end{figure}

\begin{figure}[!ht]
\begin{center}
\includegraphics[width=3.8in,angle=0]{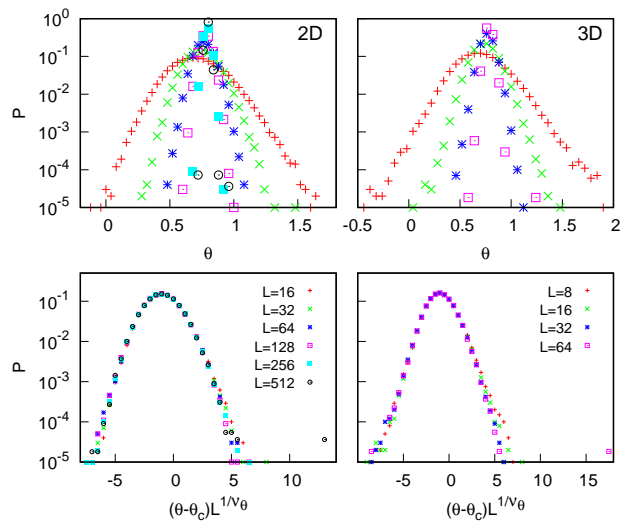}
\end{center}
\vskip -.75cm
\caption{
\label{fig_4} (Color online) \textbf{Locating the multicritical point:}
Upper panels: distribution of the location of the pseudo-multicritical points, $\theta$, for different lengths with box-$h$ randomness in 2D (left) and 3D (right). Lower panels: data collapse of the scaled distributions, with the correlation length exponent estimates from Table~\ref{table:1}. 
}
\end{figure}

Even though the QMCP is located at the classical percolation threshold, it is challenging to determine the $\theta_c$ control parameter value where the QMCP happens, due to large sample-to-sample fluctuations. It is essential to achieve this systematically, with high enough precision even at the largest sizes. 
A proven strategy is to define sample-dependent 'pseudo-multicritical' points for each sample, $s$, denoted by $\theta(s)$, by a variant of the doubling procedure\cite{2dRG}. In this doubling procedure we glue together two identical copies (also known as replicas $s$ and $s'$) of the sample by surface couplings and renormalize it up to the last site for each value of the control parameter, $\theta$. Then, for the (finite) fraction of samples that have a percolating bond dilution cluster, the renormalization is found to be qualitatively different for $\theta<\theta(s)$ and for $\theta>\theta(s)$. 
%
For weak quantum fluctuations, $\theta<\theta(s)$, the last decimated spin cluster contains equivalent sites of $s$ and $s'$. These sites and thus the two replicas are correlated and we call this cluster a \textit{correlation cluster}, see Fig.~\ref{fig_2}. The moment of the correlation cluster, $\mu(s)$, drops to zero as the pseudo-multicritical point is approached. 
This is due to the fact that for $\theta>\theta(s)$ there is no correlation cluster, i.e. there are no equivalent sites of $s$ and $s'$ in the last decimated spin cluster. $\theta(s)$ can be located iteratively with high precision in a few iterative steps. In practice, we required a precision that is at least a thousand times better than the width of the $\theta(s)$ distribution over different samples. 
Note that for samples that do not have a percolating cluster, there is no $\theta(s)$ control parameter value to result in a correlation cluster. 

With the pseudo-multicritical values at hand, we have studied the size-dependence of their distributions (in percolating samples),
which is illustrated in Figs.~\ref{fig_3} and \ref{fig_4}. Due to the broad distributions emerging from rare realizations with extreme properties, i.e. nearly disconnected percolating clusters, the mean and standard deviation of the distribution are often unreliable. As a robust estimate of the location of the multicritical points, we used the median of the distributions, $\theta(L)$, at each size, expected to scale as 
\be
|\theta_c-\theta(L)| \sim L^{-1/\nu_\theta}\;.  
\label{nu}
\ee
There are two unknowns in this formula: the true location of the multicritical point, $\theta_c$, and the $\nu_\theta$ correlation length exponent. Instead of fitting both parameters simultaneously, it is a better strategy to first get an estimate of $\nu_\theta$ from the $\Delta \theta(L) \sim L^{-1/\nu_\theta}$ scaling of the $\Delta \theta(L)$ width of the distribution. As a robust measure of the $\Delta \theta(L) $ width, we used the interquartile range (IQR). We have then calculated size-dependent effective exponents by two-point fits (comparing the results for sizes $L$ and $L/2$), which are then extrapolated as $1/L\to0$. The effective exponents are shown in the inset of Fig.~\ref{fig_3}. 
As in this example, we have generally observed that the extrapolated critical exponents are universal, i.e. independent of the form of disorder. Estimates of the extrapolated $\nu_\theta$ exponents are presented in Table~\ref{table:1}. 

With $\nu_\theta$ determined, we could now use Eq.~(\ref{nu}) to fit $\theta_c$. To obtain more precise results, following Refs.[\onlinecite{2dRG,ddRG}], we have instead formed the ratio $\alpha(L)=(\theta_c-\theta(L))/\Delta \theta(L)$. According to Eq.~(\ref{nu}), $\alpha(L)$ should be independent of the size at the multicritical point. Therefore, the true multicritical point, $\theta_c$, was determined as the value resulting in the smallest standard deviation of $\alpha(L)$:
\begin{equation*}
\begin{aligned}
\theta_c^b(\mathrm{2D})=0.783(1) \\
\theta_c^b(\mathrm{3D})=0.770(1) \\
\theta_c^f(\mathrm{2D})=-0.481(1) \\
\theta_c^f(\mathrm{3D})=-0.5055(10) 
\end{aligned}
\end{equation*}
As expected, these values are much smaller than those at the generic QCP (usually studied at $p=1$), as the correlation clusters need to overcome the dilution of the lattice. In accordance to the scaling forms used above, the resulting estimates of $\theta_c$ and $\nu_\theta$ lead to an excellent data collapse of the pseudo-multicritical point distributions as shown in Fig.~\ref{fig_4}.
Just like at the generic QCP, the value of $\alpha$ is found to be universal at the QMCP, $\alpha(\mathrm{2D})=0.43(2)$ and $\alpha(\mathrm{3D})=0.42(3)$, regardless of the form of the disorder. 

Another quantity of interest is the multicritical quantum percolation probability, $P_{q}$, which is given by the fraction of samples that have a finite replica correlation function at $\theta_c$. Our estimates are again found to be universal, $P_{q}(\mathrm{2D})=0.186(2)$ and $P_{q}(\mathrm{3D})=0.124(6)$.


\subsection{Dynamic Scaling}

\begin{figure}[!ht]
\begin{center}
\includegraphics[width=3.8in,angle=0]{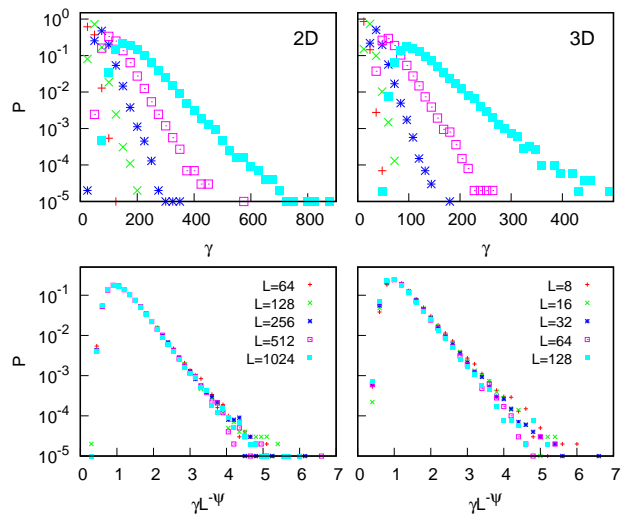}
\end{center}
\vskip -.75cm
\caption{
\label{fig_7} (Color online) \textbf{Activated dynamic scaling:}
Distribution of the log-energy parameters at the multicritical point in 2D and 3D for box-$h$ randomness (upper panels). In the lower panels the data collapse of the scaled distributions are shown, as described in the text. 
}
\end{figure}

\begin{figure}[!ht]
\begin{center}
\includegraphics[width=3.3in,angle=0]{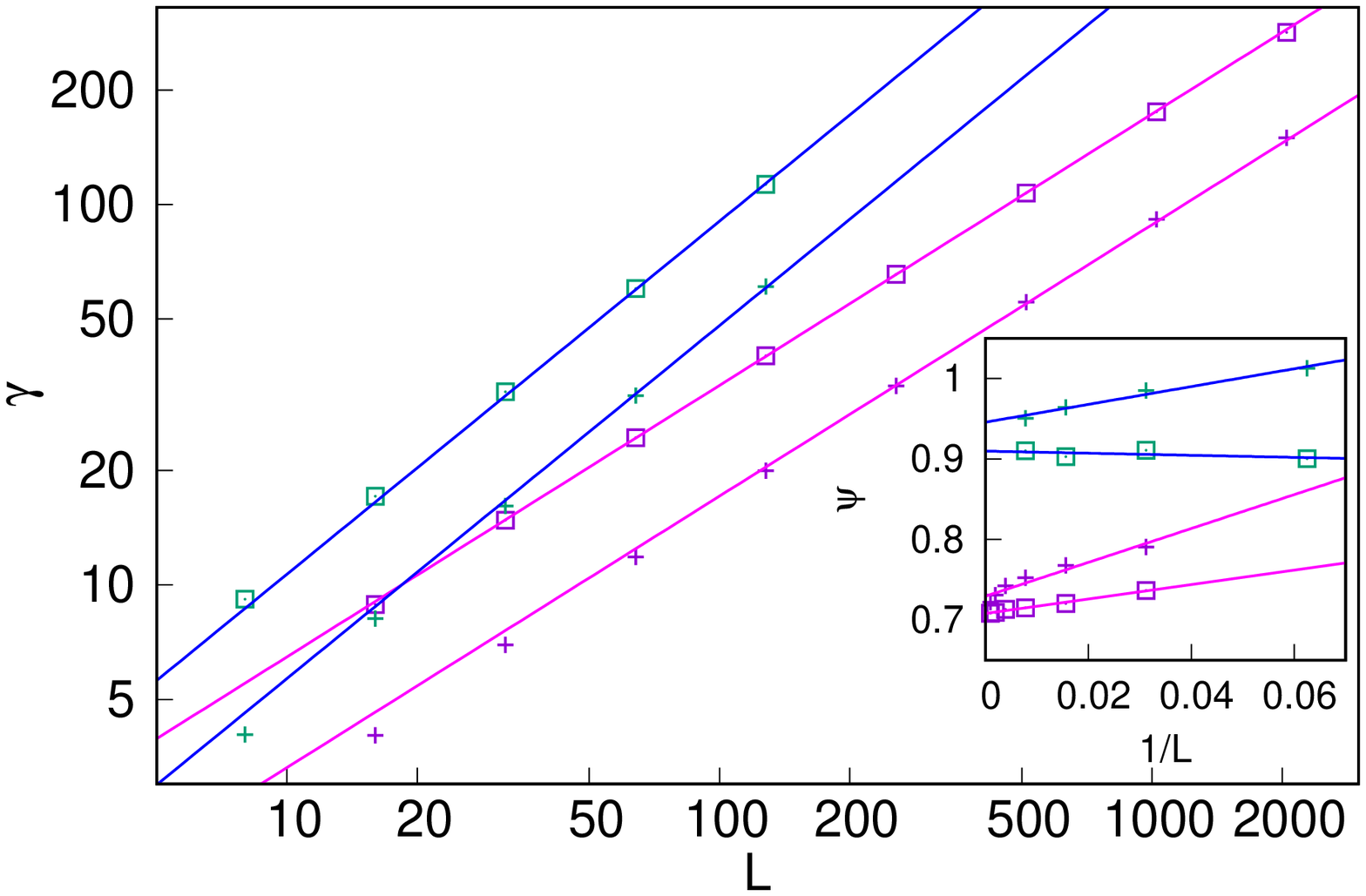}
\includegraphics[width=3.3in,angle=0]{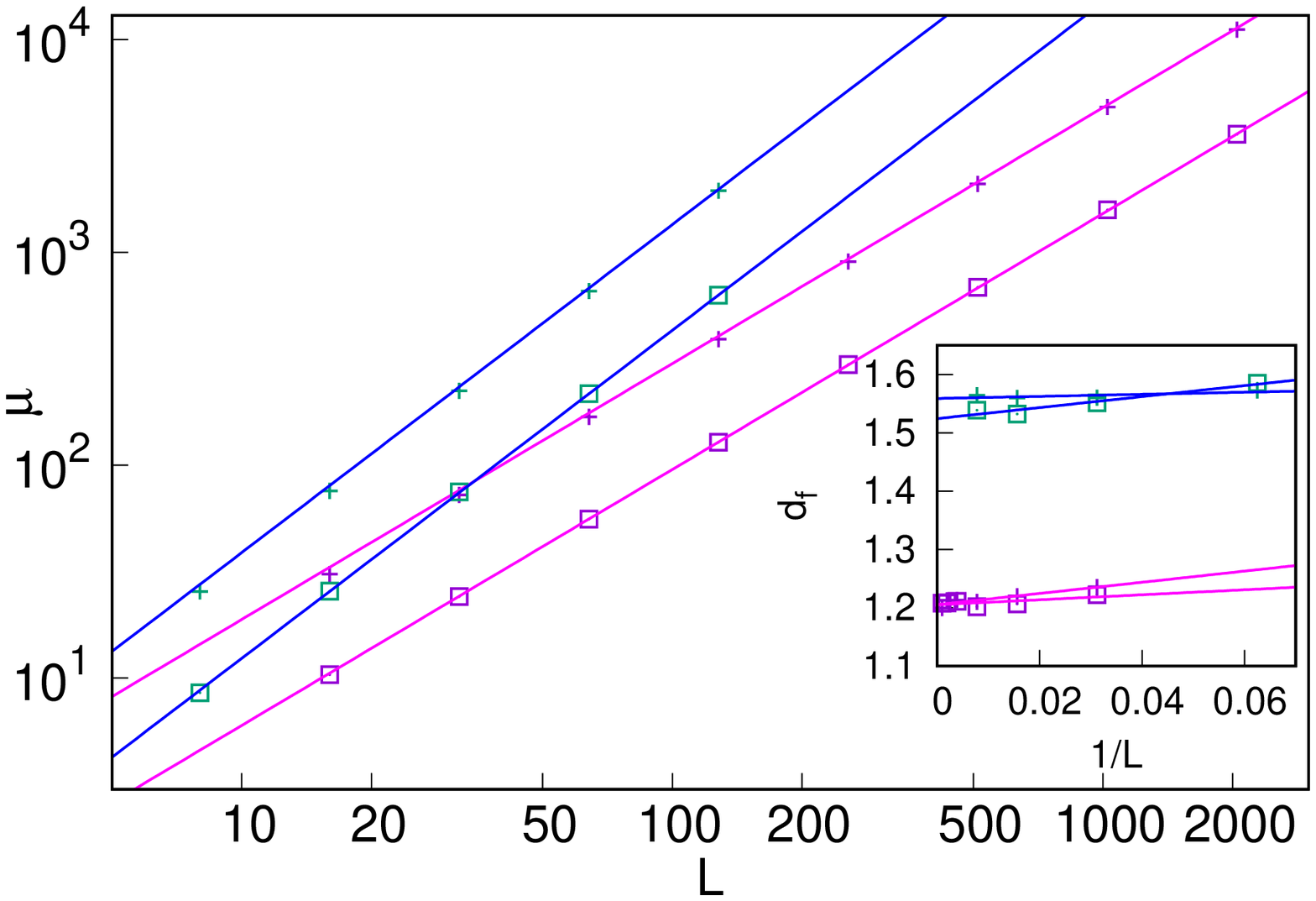}
\end{center}
\vskip -.75cm
\caption{
\label{fig_8} (Color online) \textbf{Multicritical Scaling:}
Mean log-energy parameter (top) and magnetization cluster moment (bottom) at the QMCP as the function of the length of the system in a log-log plot for the 2D (purple, magenta) and 3D (green, blue) models for two types of randomness (box-$h$: $\boxdot$, fixed-$h$: $+$). The slope of the straight lines is given by the $\psi$ (top) and $d_f$ (bottom) exponents listed in Table~\ref{table:1}. The error of the data points is smaller than the symbol size.
Insets:  Two-point estimates of $\psi$ (top) and $d_f$ (bottom) as the function of the inverse length. }
\end{figure}

Having accurate estimates for the location of the multicritical points, we have renormalized the samples at $\theta_c$ and studied the scaling behavior of the  $\mu(s)$ moment of and the $\epsilon(s)$ energy gap of the magnetization cluster.
The distribution of the log-energy parameter, $\gamma=-\ln \epsilon(s)$, is shown in the upper panels of Fig.~\ref{fig_7}, for the different models. As a clear indication of infinite disorder scaling, the width of the distribution is increasing with $L$. According to Eq.(\ref{active}), the appropriate scaling combination is $\gamma(L)L^{-\psi}$, as illustrated with the data collapse in the lower panels of Fig.~\ref{fig_7}. The critical exponent $\psi$ has been calculated from two-point fits by comparing the mean values $\gamma(L)$ and $\gamma(L/2)$ (see the upper panel of Fig.~\ref{fig_8} and Table~\ref{table:1}). 

\subsection{Fractal Dimension}

\begin{figure}[!ht]
\begin{center}
\includegraphics[width=3.8in,angle=0]{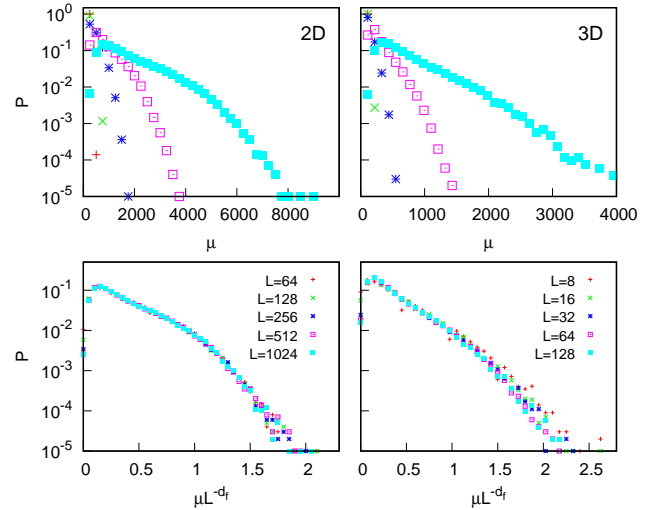}
\end{center}
\vskip -.75cm
\caption{
\label{fig_6} (Color online) \textbf{Self-similarity:}
The same as Fig.~\ref{fig_7} for the moment of magnetization clusters, see text.
}
\end{figure}

Just like at critical points, the multicritical system is expected to be statistically self-similar, i.e. free of any characteristic length scale, manifested by a non-trivial fractal dimension $d_f$. Indeed, the average moment of the magnetization cluster is found to scale as $\mu(L) \sim L^{d_f}$. We illustrate this relation in the bottom panel of Fig.~\ref{fig_8} for the 2D and the 3D models. As shown in the inset of Fig.~\ref{fig_8}, we calculated two-point fits, which are then extrapolated to $1/L\to0$, leading to the estimates in Table~\ref{table:1}. An excellent data collapse is observed for the distribution of magnetization moments for all sizes, as shown in Fig.~\ref{fig_6}. The usual magnetization is given by $m(L)\equiv\mu(L)/L^d \sim L^{-x}$, corresponding to the scaling relation $x=d-d_f$.

\subsection{Crossover Exponent}

Unlike traditional critical points, multicritical points are repulsive in multiple directions, characterized by multiple correlation length exponents, depending on the direction in which the multicritical point is crossed. So far, we have only studied small deviations in the $\theta$ control parameter at a fixed $p_c$ percolation probability, along a horizontal line through the QMCP in Fig.~\ref{fig_1}, leading to the $\nu_\theta$ exponent. 
Another possibility is to consider small deviations in $p$ while keeping $\theta_c$ fixed, along a vertical line through the QMCP in Fig.~\ref{fig_1}, leading to the  $\nu_p$ exponent. In general, small deviations in these two directions have a different impact on the correlation length. Putting it differently, finite-size effects lead to different amount of shift in the location of the multicritical point, depending on the direction in which it is crossed.
This anisotropy is then quantified by the $\phi$ crossover exponent, defined as 
\be
\left|\theta_c-\theta(L)\right|\sim\left|p_c-p(L)\right|^\phi\;,
\label{cross}
\ee
indicating $L^{-\phi/\nu_p}\sim L^{-1/\nu_\theta}$, therefore $\phi=\nu_p/\nu_\theta$.

\begin{figure}[!ht]
\begin{center}
\includegraphics[width=3.8in,angle=0]{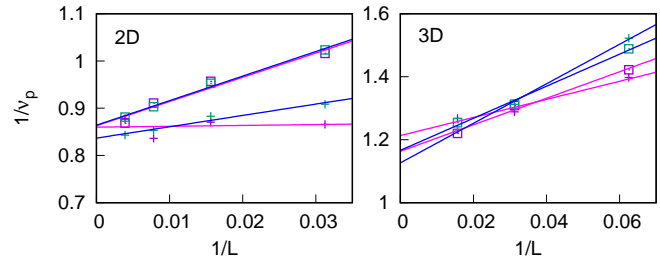}
\end{center}
\vskip -0.75cm
\caption{
\label{fig_cross} (Color online) \textbf{Varying the percolation probability:}
Two-point estimates of $1/\nu_p$ from the shift (blue) and width (magenta) of the pseudo-multicritical points as the function of the inverse length (box-$h$: $\boxdot$, fixed-$h$: $+$), when crossing the QMCP vertically in Fig.~\ref{fig_1}. Straight lines indicate the linear extrapolations, leading to the values in Table~\ref{table:1}. 
}
\end{figure}

To obtain $\nu_p$, we located the sample-dependent multicritical points in terms of $p(s)$, at the known values of $\theta_c$, similarly to the original procedure of locating the $\theta(s)$ pseudo-multicritical points at a fixed $p_c$. Note that now all $s$ samples have a finite transition in terms of $p(s)$, and it is sufficient to use the mean and standard deviation of the distributions to characterize the shift and the width. As $p_c$ is known, both the shift and the width of the distributions can be directly used to obtain finite-size estimates of $1/\nu_p$, see Fig.~\ref{fig_cross}. Interestingly, at the 3D QMCP, $\nu_p$ is within the error the same as at the percolation QCP. In all cases, the extrapolated $\nu_p$ exponents are much smaller than the $\nu_\theta$ exponents, leading to $\phi<1$, see Table~\ref{table:1}. We have also obtained estimates for the $\phi$ crossover exponent directly from Eq.~(\ref{cross}), leading to the same values as the ones calculated as the ratio of correlation length exponents.


\section{Discussion}
\label{Sec:discussion}



Our SDRG results indicate that the quantum multicritical behavior of the random transverse-field Ising model in two and three dimensions shows ultraslow, activated dynamic scaling, controlled by infinite disorder fixed points. This observation justifies the use of the SDRG method, by ensuring that the calculated numerical results get asymptotically correct as the system size increases. The multicritical exponents presented in Table~\ref{table:1} are found to be universal, i.e. independent from the type of ferromagnetic disorder. 

As expected, the $d_f$ fractal dimension and the $\psi$ dynamical exponents are found to interpolate between those corresponding to the percolation and generic QCPs. 
%
%
As a guide for future experiments at QMCPs, singularities of the thermodynamic quantities at small temperatures involve the exponents in Table~\ref{table:1}.
For example, the susceptibility and the specific heat behave as $\chi(T) \sim |\log T|^{(2d_f-d)/\psi}/T$ and
$C_V(T) \sim |\log T|^{d/\psi}$, respectively\cite{danielreview,im}. 

\begin{table}[ht]
\caption{\textbf{Critical and multicritical exponents of the RTIM:} The number in the parentheses gives an error estimate of the last digits. CP stands for the multicritical point of the contact process. '$-$' indicates an exponent of unknown value.
 \label{table:1}}
 \begin{tabular}{cc|c|c|c|c|c|c|}  
  & & Percolation&Generic& CP& QMCP\\ 
    & & QCP [\onlinecite{stauffer, xu}]& QCP [\onlinecite{2dRG,ddRG}] & [\onlinecite{dahmen}] & this work \\ \hline
2D&$\psi$    & $91/48\sim1.896$  & $ 0.48(2)$ & $0.57(4)$ &$0.708(20)$ \\ 
&$d_f$    & $91/48\sim1.896$ & $1.018(15)$ & $1.02(18)^{\dagger}$ &$1.205(3)$ \\ 
&$\nu_\theta$    &  & $1.24(2)$ & $0.88(10)^* $ &$1.382(7)$  \\ 
&$\nu_p$    & $4/3\sim1.333$ &  & $-$ &$1.168(10)$  \\ 
&$\phi$    &  &  & $-$ &$0.845(11)$  \\  \hline
3D&$\psi$    & $2.52293(10)$  & $ 0.46(2)$ & $-$ &$0.93(2)$\\ 
&$d_f$    & $2.52293(10)$ & $1.160(15)$ & $-$ &$1.54(2)$  \\ 
&$\nu_\theta$    &   & $0.98(2)$ & $-$ &$1.123(10)$\\ 
&$\nu_p$    & $0.8762(12)$  & & $-$ &$0.86(1)$\\ 
&$\phi$    &   & & $-$ &$0.76(2)$\\ \hline
  \end{tabular}
  \end{table}

For sufficiently strong initial disorder, the SDRG critical exponents are known to be in agreement with the Monte Carlo results on the 2D and 3D random contact process\cite{vojta09, vojta3D, hiv}, which is a simple nonequilibrium model of infection spreading. As the QCPs are in the same universality classes, it is natural to expect that the multicritical random contact process is in the same universality class as the QMCP of the RTIM, at least for strong enough disorder. The corresponding multicritical point of the random contact process has been studied in 2D, although with considerable difficulties, leading to admittedly large systematic errors\cite{dahmen}. Nevertheless, clear signs of activated dynamic scaling have been observed, in agreement to our results. In addition, the cluster of active sites at long times in Fig.~2 of Ref.[\onlinecite{dahmen}] appears to be qualitatively similar to our Fig.~\ref{fig_2}, obtained by the SDRG method.
At the quantitative level, however, these early simulations might to be less reliable, as the reported correlation length exponent, $\nu=0.88(10)$ (marked by an asterisk in Table~\ref{table:1}), fails the celebrated bound, known as the Harris-criterion\cite{ccfs}: $\nu \ge 2/d=1$. 
While there are known cases where the Harris-criterion is violated\cite{barghathi,violation}, it would require careful investigations to determine if this is the case for the multicritical random contact process. In addition to providing the first reliable values for $\nu_\theta$ in 2D, our results present the first values for both $\nu_p$ and the $\phi$ crossover exponents. In both 2D and 3D, our $\nu_p$ and $\nu_\theta$ correlation length exponents satisfy the Harris-criterion.

Note that, when comparing further exponents with those reported in Ref.[\onlinecite{dahmen}], we need to compensate for the fact that the density of the clusters was measured w.r.t. the largest cluster only, instead of the full system. Therefore, the reported $x'\equiv\beta/\nu=0.81/0.88=0.92$ exponent needs to be adjusted to $x=x'+5/91=0.98(18)$, leading to a shift in $d_f= d-x$, marked by $\dagger$ in Table~\ref{table:1}. This adjusted value of $d_f$ is in agreement to our results. 
In 3D, to the best of our knowledge, we present the first results at the QMCP of the RTIM, providing valuable guidance for future quantum experiments as well as for Monte Carlo studies of the disordered directed percolation universality class, including the random contact process.

%
%

Finally, we note that our SDRG investigations can be extended in several directions. Here, we mention the characterization of the entanglement entropy\cite{refael,lin07,yu07, kovacs, kovacs3dperc, kovacs_igloi12}, transverse correlations\cite{transverse}, boundary critical exponents\cite{boundary}, the impact of long-range interactions\cite{long-range3d, long-rangeCP}, as well as the dynamical singularities in the disordered and ordered Griffiths phases\cite{im, 2dRG, ddRG}.

\begin{acknowledgments}
We are grateful to F. Igl\'oi for useful discussions.

\end{acknowledgments}

\end{document}